\documentclass[aip,apl,10pt,superscriptaddress,twocolumn,reprint]{revtex4-1}

\usepackage{graphicx}
\usepackage{bpchem}
\usepackage{amssymb}
\usepackage{color}

\begin{document}

\title{Tuning the giant inverse magnetocaloric effect in \BPChem{Mn\_{2-x}Cr\_{x}Sb} compounds}

\author{L. Caron} 
\email[]{L.Caron@tudelft.nl}
\affiliation{Fundamental Aspects of Materials and Energy, Faculty of Applied Sciences, TUDelft, Mekelweg 15, 2629 JB Delft, The Netherlands}
\affiliation{Instituto de Física Gleb Wataghin, Universidade Estadual de Campinas-UNICAMP, C.P. 6165, Campinas 13 083 970, SP, Brazil}
\author{X.F. Miao}
\affiliation{Fundamental Aspects of Materials and Energy, Faculty of Applied Sciences, TUDelft, Mekelweg 15, 2629 JB Delft, The Netherlands}
\author{J. C. P Klaasse}
\affiliation{Van der Waals-Zeeman Instituut, Universiteit van Amsterdam, Science Park 904
1098 XH Amsterdam, The Netherlands}
\author{S. Gama}
\affiliation{Departamento de Ci\^{e}ncias Exatas e da Terra, Universidade Estadual de S\~{a}o Paulo - Unifesp, 09972-270, Diadema, Brazil}
\affiliation{Instituto de Física Gleb Wataghin, Universidade Estadual de Campinas-UNICAMP, C.P. 6165, Campinas 13 083 970, SP, Brazil}
\author{E. Br\"{u}ck}
\affiliation{Fundamental Aspects of Materials and Energy, Faculty of Applied Sciences, TUDelft, Mekelweg 15, 2629 JB Delft, The Netherlands}
\date{\today}

\begin{abstract}
Structural, magnetic and magnetocaloric properties of \BPChem{Mn\_{2-x}Cr\_{x}Sb} compounds have been studied. In these compounds a first order magnetic phase transition from the ferrimagnetic to the antiferromagnetic state occurs with decreasing temperature, giving rise to giant inverse magnetocaloric effects that can be tuned over a wide temperature interval through changes in substitution concentration.
Entropy changes as high as 7.5 J/kgK have been observed, and a composition independent entropy change is obtained for several different concentrations/working temperatures, making these compounds suitable candidates for a composite working material.

\end{abstract}
\maketitle
\section{\label{intro}Introduction}
In the past few decades, with the discovery and development of materials presenting first order magneto-structural transitions around room-temperature, refrigeration based on the magnetocaloric effect (MCE) has become an environmentally friendly and efficient alternative to gas-compression-based refrigeration.

The MCE presents itself as the magnetic entropy and adiabatic temperature changes observed when a magnetic material is submitted to an external magnetic field change. While it is intrinsic to all magnetic materials, it is maximal around magnetic phase transitions, particularly coupled  magneto-structural first-order phase transitions. Most materials presenting a large magnetocaloric effect rely on order-disorder type phase transitions coupled to discontinuous structural changes. Materials such as MnAs\cite{wada:3302},  \BPChem{Gd\_{5}(Si\_{2}Ge\_{2})}\cite{GdGeSi} and MnCoGe-based compounds\cite{trung:162507, trung:172504, caron_pressure-tuned_2011} present crystal symmetry changes with large volume variations. On the other hand, the most promising materials for applications do not, showing only a discontinuity in lattice parameters with smaller or no volume changes at all, as is the case for \BPChem{La(Fe\_{1-x}Si\_{x})\_{13}}\cite{Fujita2003} and \BPChem{(Mn,Fe)\_{2}(P,Si)}\cite{Dung-Mixed}, respectively.

However, large entropy changes can be originated not only from order-disorder transitions but also from a number of order-order first-order phase transitions, such as spin flip, spin reorientation, antiferro to ferro or ferrimagnetic phase transitions.

\BPChem{Mn\_{2}Sb}\cite{wilkinson_magnetic_1957} is a ferrimagnet with \BPChem{T\_{C}} around 550 K and tetragonal \BPChem{Cu\_{2}Sb}-type structure (space group \BPChem{D\^{7}\_{4h}}-P4/nmm). In this system the magnetism is solely due to the Mn atoms, which occupy two non-equivalent crystallographic sites \textsl{2a} and \textsl{2c}, hereon referred as \BPChem{Mn\_{I}} and \BPChem{Mn\_{II}}, respectively. The magnetic structure, as described by Cloud et al\cite{cloud_neutron_1960} , consists of \BPChem{Mn\_{I}} and \BPChem{Mn\_{II}} sublattices stacked in triple layers in a \BPChem{Mn\_{I}}-\BPChem{Mn\_{II}}-\BPChem{Mn\_{II}} structure which repeats along the c-direction. Both sublattices show ferromagnetic intralayer interactions, with the \BPChem{Mn\_{II}} moment being roughly twice that of \BPChem{Mn\_{I}}, both parallel to the tetragonal axis. However, adjacent \BPChem{Mn\_{I}}-\BPChem{Mn\_{II}} layers couple antiparallel while adjacent \BPChem{Mn\_{II}}-\BPChem{Mn\_{II}} layers couple parallel to each other, resulting in a ferrimagnetic \BPChem{Mn\_{I}}-\BPChem{Mn\_{II}}-\BPChem{Mn\_{II}} $\upharpoonleft \downharpoonright \downharpoonright$ configuration.

It is well known that for Mn compounds the exchange interaction is strongly dependent on the interatomic spacing. Therefore, below a critical distance between adjacent \BPChem{Mn\_{II}} layers, due to the Pauli exclusion principle, the interlayer exchange interactions must change sign and become antiferromagnetic with a $\upharpoonleft \downharpoonright \upharpoonleft$ configuration, in a phenomenon known as exchange inversion\cite{swoboda_evidence_1960}. The transition from the ferri to the antiferromagnetic state is accompanied by the rotation of the moments from the direction parallel to that perpendicular to the tetragonal axis. As this translates into a decrease of the magnetization during lattice contraction or cooling a so-called inverse transition is observed. 

For pure \BPChem{Mn\_{2}Sb} normal thermal contraction is not sufficient to trigger exchange inversion, but if a substitution is introduced which decreases the lattice parameters, the critical distance at which exchange inversion takes place may become accessible.
Exchange inversion can be achieved by a number of different substitutions both in the Mn as in the Sb sites. They can be smaller atoms which contract the lattice: Cr, Co, V and Cu on the Mn site, and As, Ge and Sn on the Sb site\cite{bither_new_1962, flippen_entropy_1963, kanomata_magnetic_1984, zhang_metamagnetic-transition-induced_2003}, or a larger atom which drastically changes the compressibility of the lattice, as is the case of Bi substitution on the Sb site\cite{ohshima_magnetic_1979}. As the interlayer exchange interaction changes sign, the variation of the \textit{c} lattice parameter becomes critical. At the transition a large discontinuous change of the \textit{c} parameter is observed, which is compensated by an opposite variation of the \textit{a} parameter, resulting in small volume changes. 

How much the c parameter is decreased can be controlled through the amount of substitution, making it possible to tune the temperature at which the exchange inversion occurs and therefore the working temperature of a given compound.

In this paper we report on the properties of the \BPChem{Mn\_{2-x}Cr\_{x}Sb} series of compounds which present the so-called inverse magnetocaloric effect, associated with a first-order phase transition between antiferro and ferrimagnetic states.

\section{\label{exp}Experimental Methods}
Polycrystalline \BPChem{Mn\_{2-x}Cr\_{x}Sb} was prepared from high purity elements: Mn and Sb in pieces and Cr in flakes. These samples were carefully molten in an arc-melting furnace under high purity Ar atmosphere. The buttons were molten from 5 to 6 times and turned upside down in between melts to ensure homogeneity. Weight losses as low as 0.5\% were observed.
The buttons obtained by arc-melting were then sealed in quartz ampoules under Ar atmosphere and annealed for 120 h at 1073 K. Heating and cooling to and from the annealing temperature was performed at a rate of 5 K/min.

The samples were structurally characterized at room temperature through X-ray diffraction measurements using Cu-\BPChem{K$\alpha$\_{1}} radiation which were analyzed using \textsc{fullprof}'s implementation of the Rietveld refinement method \cite{FullProf,rietveld}. 

The magnetic and magnetocaloric properties around the antiferro-ferrimagnetic transition were measured on a Quantum Design MPMS 5 SQUID magnetometer. The Curie temperatures were measured in a Lake Shore VSM.
Other isofield magnetization measurements were performed in a Quantum Design PPMS. The magnetic entropy change was calculated from isothermal magnetization data using the Maxwell relations\cite{caron_2009}.

Calorimetry measurements were performed using the setup built and reported by Klaasse et al.\cite{klaasse_heat-capacity_2008} as well as a commercial TA instruments Q2000 DSC.

\section{Results and Discussion}
Four compositions of \BPChem{Mn\_{2-x}Cr\_{x}Sb} were studied x = 0.06; 0.08; 0.10 and 0.12, with transition temperatures within the range from 220 K to 340 K. 

\begin{figure}
\includegraphics[scale=0.85]{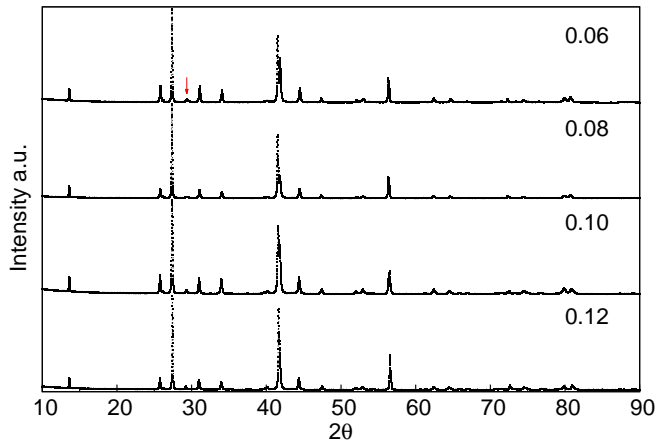}
\caption{\label{XRD}X-ray diffraction patterns at room temperature for the \BPChem{Mn\_{2-x}Cr\_{x}Sb} compounds. The red arrow identifies the reflection attributed to the MnSb secondary phase.}
\end{figure}

\begin{figure}
\includegraphics[scale=0.85]{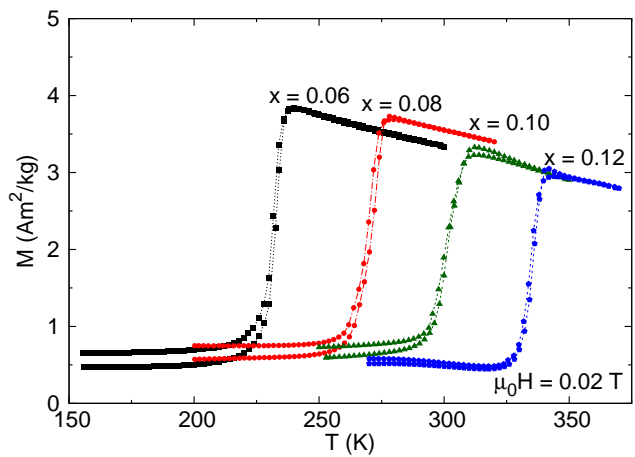}
\caption{\label{MxT}Temperature dependence of the magnetization.}
\end{figure}

XRD at room temperature shows that all four samples crystallize in the tetragonal \BPChem{Cu\_{2}Sb}-type of structure (see figure \ref{XRD}). The lattice parameters obtained from Rietveld refinement are presented in table \ref{table-lattice}. Increasing Cr content increases the \textit{a} and decreases the \textit{c} lattice parameter. For x $\geq$ 0.10 \textit{c} is below the critical distance \BPChem{\textit{c}\_{crit} $\simeq$ 6.53 \AA} for wich the antiferromagnetic phase is accessible\cite{swoboda_evidence_1960} in \BPChem{Mn\_{2-x}Cr\_{x}Sb} compounds, and therefore the transition temperature \BPChem{T\_{t}} is above room temperature (see figure \ref{MxT}). The weak reflection marked on figure \ref{XRD} is attributed to a small amount of second phase, identified as MnSb precipitate which crystallizes in the hexagonal \BPChem{Ni\_{2}In}-type of structure (space group \BPChem{D\_{6h}\^{4}}-\BPChem{P 6\_{3}/mmc}). This ferromagnetic impurity has a \BPChem{T\_{C}} around 585 K and is responsible for the rather high magnetic response in the antiferromagnetic state\cite{okita_crystal_1968} (see figure \ref{MxT}).

\begin{figure}
\includegraphics[scale=0.85]{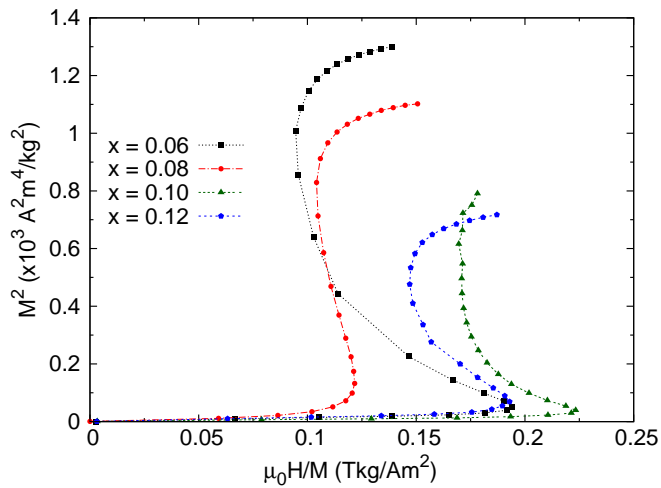}\\[1.5ex]
\caption{\label{arrott}Arrott plots around the phase transition temperature for all compositions.}
\end{figure}

\begin{table}[ht]
\setlength{\tabcolsep}{6pt}
\caption{Lattice parameters and volume derived from room temperature XRD measurements.}
\begin{centering}
\begin{tabular}{c c c c}
\hline\hline
x &   a ({\AA}) &   c ({\AA}) &    Vol (\BPChem{{\AA}\^{3}}) \\[0.5ex]
\hline
0.06 &  4.07952(3) &  6.53651(3) &  108.784(1) \\
0.08 &  4.08075(3) &  6.53330(2) &  108.796(1) \\
0.10 &  4.08708(3) &  6.52545(3) &  109.003(1) \\
0.12 &  4.08859(3) &  6.51117(2) &  108.844(1) \\[1ex]
\hline
\end{tabular}
\end{centering}
\label{table-lattice}
\end{table}
Isofield magnetization measurements (see figure \ref{MxT}) and Arrott plots (see figure \ref{arrott}) reveal very sharp first-order phase transitions with very low thermal hysteresis (less than 2 K for all compositions reported).  The transition temperature from the antiferro to the ferrimagnetic state \BPChem{T\_{t}} is very sensitive to composition and increases with increasing Cr content at a rate of 8.2 K/at.\% Cr. 
Unlike \BPChem{T\_{t}} which quickly increases with increasing Cr substitution, the transition temperature between the ferri and paramagnetic states \BPChem{T\_{C}} slowly decreases with increasing Cr content at a rate -2.3 K/at.\% Cr. Therefore, increasing Cr substitution effectively brings the antiferro to ferri and the ferri to paramagnetic transitions together. However, for all compositions studied, \BPChem{T\_{C}} remains far above \BPChem{T\_{t}} and does not hinder the use of the inverse transition in applications. 

\begin{figure}
\includegraphics[scale=0.85]{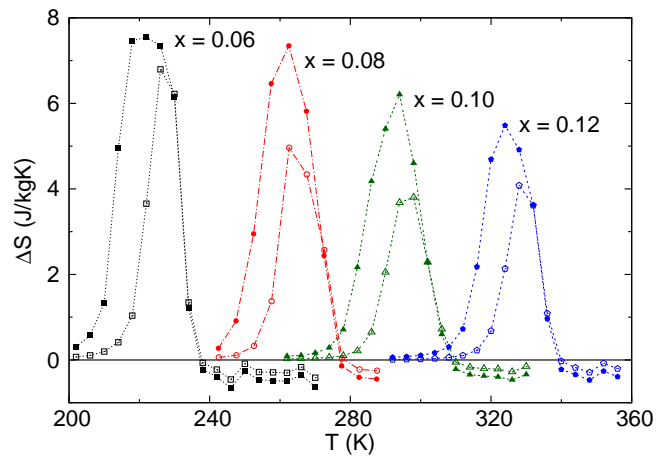}
\caption{\label{dSxT}Temperature dependence of the entropy change for a 2 T (open symbols) and 5 T (closed symbols) magnetic field changes.}
\end{figure}

The thermodynamical quantities are summarized in table \ref{table-thermo}. The saturation magnetization (right above the transition temperature at 5 T) decreases linearly with increasing Cr content, as the amount of Mn is decreased. This decrease reflects directly on the magnitude of the entropy change (see figure \ref{dSxT} and table \ref{table-thermo}), which also decreases with increasing Cr content.

Notice that, unlike most magnetocaloric materials, the entropy change due to the variation of the external applied magnetic field is positive for \BPChem{Mn\_{2-x}Cr\_{x}Sb} compounds, reflecting the fact that the magnetization increases with increasing temperature through the antiferro to ferrimagnetic phase transition. Also, since magnetic field stabilizes the ferrimagnetic state, the \BPChem{T\_{t}} shifts to lower temperatures with increasing magnetic field.

Entropy changes are found to be composition independent over the composition interval studied, ranging from 7.5 J/kgK to 5.5 J/kgK for a magnetic field change from 0 to 5 T. This means that using different compositions of \BPChem{Mn\_{2-x}Cr\_{x}Sb} it is possible to produce a composite material presenting a table-like effect (i.e., a plateau in the \BPChem{$\Delta$S} curve) over a temperature interval of approximately 100 K. Applications for such composite material could be found where cooling is required over large temperature spans in spite of a lower cooling power. 
 
\begin{figure}
\includegraphics[scale=0.85]{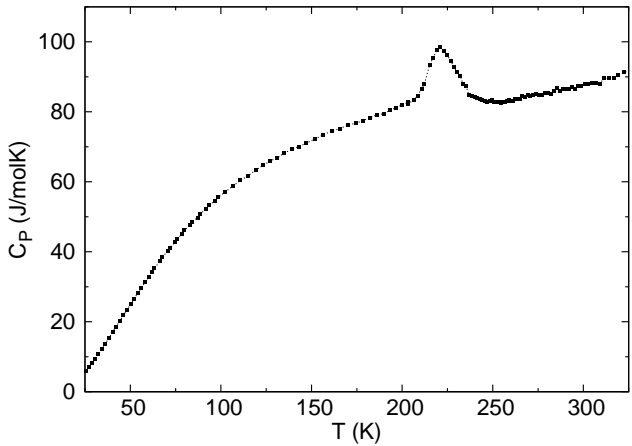}
\caption{\label{CpxT}Heat capacity as a function of temperature for the sample x = 0.06.}
\end{figure}

\begin{table}[ht]
\setlength{\tabcolsep}{3pt}
\caption{Thermodynamic quantities.}
\centering
\begin{tabular}{c c c c c c c c}
\hline\hline
x & \BPChem{T\_{t}} & \BPChem{T\_{C}} & \BPChem{$\Delta$S\_{M}} & \BPChem{dT\_{t}/d$\mu$\_{0}H}  & \BPChem{M\_{S}} & \BPChem{$\Delta$T\_{ad}\^{max}} \\
  & \small{K} & \small{K} & \small{J/kgK} & \small{K/T} & \small{\BPChem{Am\^{2}/kg}} & \small{K} \\
 & \footnotesize{0.02 T} & \footnotesize{0.02 T} & \footnotesize{0 - 5 T} & & \footnotesize{5 T} & \footnotesize{0 - 1 T} &\\[0.5ex]
\hline
0.06 & 232 & 520 & 7.5 & - 3.9 & 36 & 6.1 &  \\
0.08 & 272 & 510 & 7.3 & - 4.3 & 33 & 5.9 & \\
0.10 & 300 & 505 & 6.2 & - 3.7 & 29 & 6.7 & \\
0.12 & 333 & 490 & 5.5 & - 4.1 & 27 & 6.2 & \\[1ex]
\hline
\end{tabular}
\label{table-thermo}
\end{table}
The specific heat was measured for the sample with x = 0.06 and is presented in figure \ref{CpxT}. The latent heat content of the peak (minus an ``eye-ball''-shaped background) is 275 J/mol, and the entropy change due to the transition in the absence of an external magnetic field can be estimated to be approximately 5.1 J/kgK. The difference with the value in table \ref{table-thermo} can be attributed to the reported systematic errors made by the step wise methods in determining the latent heat of first order transitions, resulting in too low values\cite{ADMA:ADMA200901435}. Magnetic field shifts the first order phase transition to lower temperatures widening the entropy change peak (and thus increasing the cooling power) but does not significantly increase it's maximum.

Another important feature for applications is the rate at which the transition temperature shifts due to the external magnetic field, \BPChem{$\delta$T\_{t}/$\delta$B} (see table \ref{table-thermo}). As a magnetocaloric material is cycled in and out of a magnetic field, the reversibility of the cycle will depend on the thermal hysteresis as well as \BPChem{$\delta$T\_{t}/$\delta$B}. The \BPChem{Mn\_{2-x}Cr\_{x}Sb} compounds show low thermal hysteresis and large \BPChem{$\delta$T\_{t}/$\delta$B} being ideal for applications. Moreover the first order ferrimagnetic to antiferromagnetic phase transition is accompanied by discontinuous jumps in both \textit{a} and \textit{c} lattice parameters in opposite senses, resulting in volume changes from 0.1\% to 0.02\% in the studied range\cite{}. In first order phase transitions where discontinuities in the thermal evolution of the crystal lattice are intrinsic, small volume changes are critical in keeping physical stability during cycling.

This shift is not only important in determining the operating conditions for applications but it also provides an upper bound to the value of the adiabatic temperature change \BPChem{$\Delta$T\_{ad}\^{max}}\cite{sandeman_magnetocaloric_2012}. The upper bound to \BPChem{$\Delta$T\_{ad}} can be calculated using \BPChem{$\delta$T\_{t}/$\delta$B}, \BPChem{M\_{S}} and the specific heat value just before (or after) the latent heat peak of the first order phase transition (see figure \ref{CpxT}). For the studied compounds the upper bound is found at \BPChem{$\approx$ 6 K} in a field change of 1 T (see table \ref{table-thermo}), higher values than found for materials such as \BPChem{La(Fe\_{1-x}Si\_{x})\_{13}}\cite{Fujita2003} and \BPChem{(Mn,Fe)\_{2}(P,Si)}\cite{Dung-Mixed}.

\section{Conclusions}
In summary we have studied the \BPChem{Mn\_{2-x}Cr\_{x}Sb} compounds showing a first order magnetic phase transition between antiferro and ferrimagnetic states which can be tuned by substitutions. The order-order transition gives rise to moderate inverse magnetocaloric effect which is found to be composition-independent within a large range of compositions and temperatures. Together with small volume changes and large \BPChem{$\delta$T\_{t}/$\delta$B} and \BPChem{$\Delta$T\_{ad}\^{max}}, the composition-independent entropy changes make these compounds ideal for applications where heat pumping over large temperature ranges is necessary.\\

\begin{acknowledgments}
The authors acknowledge Funda\c{c}\~{a}o de Amparo \`{a} Pesquisa do Estado de S\~{a}o Paulo - FAPESP(03/12604-6) and Coordena\c{c}\~{a}o de Aperfei\c{c}oamento de Pessoal de N\'{i}vel Superior - CAPES (BEX 4631-06-4) for financial support. This work is also part of an Industrial Partnership Programme IPP I28 of the ``Stichting voor Fundamenteel Onderzoek der Materie (FOM)'' which is financially supported by the ``Nederlandse Organisatie voor Wetenschappelijk Onderzoek (NWO)'' and co-financed by BASF Future Business.
\end{acknowledgments}

\begin{thebibliography}{22}%
\makeatletter
\providecommand \@ifxundefined [1]{%
 \@ifx{#1\undefined}
}%
\providecommand \@ifnum [1]{%
 \ifnum #1\expandafter \@firstoftwo
 \else \expandafter \@secondoftwo
 \fi
}%
\providecommand \@ifx [1]{%
 \ifx #1\expandafter \@firstoftwo
 \else \expandafter \@secondoftwo
 \fi
}%
\providecommand \natexlab [1]{#1}%
\providecommand \enquote  [1]{``#1''}%
\providecommand \bibnamefont  [1]{#1}%
\providecommand \bibfnamefont [1]{#1}%
\providecommand \citenamefont [1]{#1}%
\providecommand \href@noop [0]{\@secondoftwo}%
\providecommand \href [0]{\begingroup \@sanitize@url \@href}%
\providecommand \@href[1]{\@@startlink{#1}\@@href}%
\providecommand \@@href[1]{\endgroup#1\@@endlink}%
\providecommand \@sanitize@url [0]{\catcode `\\12\catcode `\$12\catcode
  `\&12\catcode `\#12\catcode `\^12\catcode `\_12\catcode `\%12\relax}%
\providecommand \@@startlink[1]{}%
\providecommand \@@endlink[0]{}%
\providecommand \url  [0]{\begingroup\@sanitize@url \@url }%
\providecommand \@url [1]{\endgroup\@href {#1}{\urlprefix }}%
\providecommand \urlprefix  [0]{URL }%
\providecommand \Eprint [0]{\href }%
\providecommand \doibase [0]{http://dx.doi.org/}%
\providecommand \selectlanguage [0]{\@gobble}%
\providecommand \bibinfo  [0]{\@secondoftwo}%
\providecommand \bibfield  [0]{\@secondoftwo}%
\providecommand \translation [1]{[#1]}%
\providecommand \BibitemOpen [0]{}%
\providecommand \bibitemStop [0]{}%
\providecommand \bibitemNoStop [0]{.\EOS\space}%
\providecommand \EOS [0]{\spacefactor3000\relax}%
\providecommand \BibitemShut  [1]{\csname bibitem#1\endcsname}%
\let\auto@bib@innerbib\@empty
\bibitem [{\citenamefont {Wada}\ and\ \citenamefont
  {Tanabe}(2001)}]{wada:3302}%
  \BibitemOpen
  \bibfield  {author} {\bibinfo {author} {\bibfnamefont {H.}~\bibnamefont
  {Wada}}\ and\ \bibinfo {author} {\bibfnamefont {Y.}~\bibnamefont {Tanabe}},\
  }\href@noop {} {\bibfield  {journal} {\bibinfo  {journal} {Appl. Phys.
  Lett.}\ }\textbf {\bibinfo {volume} {79}},\ \bibinfo {pages} {3302} (\bibinfo
  {year} {2001})}\BibitemShut {NoStop}%
\bibitem [{\citenamefont {Pecharsky}\ and\ \citenamefont
  {Gschneidner}(1997)}]{GdGeSi}%
  \BibitemOpen
  \bibfield  {author} {\bibinfo {author} {\bibfnamefont {V.~K.}\ \bibnamefont
  {Pecharsky}}\ and\ \bibinfo {author} {\bibfnamefont {K.~A.}\ \bibnamefont
  {Gschneidner}, \bibfnamefont {Jr.}},\ }\href {\doibase
  10.1103/PhysRevLett.78.4494} {\bibfield  {journal} {\bibinfo  {journal}
  {Phys. Rev. Lett.}\ }\textbf {\bibinfo {volume} {78}},\ \bibinfo {pages}
  {4494} (\bibinfo {year} {1997})}\BibitemShut {NoStop}%
\bibitem [{\citenamefont {Trung}\ \emph
  {et~al.}(2010{\natexlab{a}})\citenamefont {Trung}, \citenamefont {Biharie},
  \citenamefont {Zhang}, \citenamefont {Caron}, \citenamefont {Buschow},\ and\
  \citenamefont {Br\"uck}}]{trung:162507}%
  \BibitemOpen
  \bibfield  {author} {\bibinfo {author} {\bibfnamefont {N.~T.}\ \bibnamefont
  {Trung}}, \bibinfo {author} {\bibfnamefont {V.}~\bibnamefont {Biharie}},
  \bibinfo {author} {\bibfnamefont {L.}~\bibnamefont {Zhang}}, \bibinfo
  {author} {\bibfnamefont {L.}~\bibnamefont {Caron}}, \bibinfo {author}
  {\bibfnamefont {K.~H.~J.}\ \bibnamefont {Buschow}}, \ and\ \bibinfo {author}
  {\bibfnamefont {E.}~\bibnamefont {Br\"uck}},\ }\href@noop {} {\bibfield
  {journal} {\bibinfo  {journal} {Appl. Phys. Lett.}\ }\textbf {\bibinfo
  {volume} {96}},\ \bibinfo {pages} {162507} (\bibinfo {year}
  {2010}{\natexlab{a}})}\BibitemShut {NoStop}%
\bibitem [{\citenamefont {Trung}\ \emph
  {et~al.}(2010{\natexlab{b}})\citenamefont {Trung}, \citenamefont {Zhang},
  \citenamefont {Caron}, \citenamefont {Buschow},\ and\ \citenamefont
  {Br\"uck}}]{trung:172504}%
  \BibitemOpen
  \bibfield  {author} {\bibinfo {author} {\bibfnamefont {N.~T.}\ \bibnamefont
  {Trung}}, \bibinfo {author} {\bibfnamefont {L.}~\bibnamefont {Zhang}},
  \bibinfo {author} {\bibfnamefont {L.}~\bibnamefont {Caron}}, \bibinfo
  {author} {\bibfnamefont {K.~H.~J.}\ \bibnamefont {Buschow}}, \ and\ \bibinfo
  {author} {\bibfnamefont {E.}~\bibnamefont {Br\"uck}},\ }\href {\doibase
  10.1063/1.3399773} {\bibfield  {journal} {\bibinfo  {journal} {Appl. Phys.
  Lett.}\ }\textbf {\bibinfo {volume} {96}},\ \bibinfo {pages} {172504}
  (\bibinfo {year} {2010}{\natexlab{b}})}\BibitemShut {NoStop}%
\bibitem [{\citenamefont {Caron}, \citenamefont {Trung},\ and\ \citenamefont
  {Br\"uck}(2011)}]{caron_pressure-tuned_2011}%
  \BibitemOpen
  \bibfield  {author} {\bibinfo {author} {\bibfnamefont {L.}~\bibnamefont
  {Caron}}, \bibinfo {author} {\bibfnamefont {N.~T.}\ \bibnamefont {Trung}}, \
  and\ \bibinfo {author} {\bibfnamefont {E.}~\bibnamefont {Br\"uck}},\ }\href
  {\doibase 10.1103/PhysRevB.84.020414} {\bibfield  {journal} {\bibinfo
  {journal} {Physical Review B}\ }\textbf {\bibinfo {volume} {84}},\ \bibinfo
  {pages} {020414} (\bibinfo {year} {2011})}\BibitemShut {NoStop}%
\bibitem [{\citenamefont {Fujita}\ \emph {et~al.}(2003)\citenamefont {Fujita},
  \citenamefont {Fujieda}, \citenamefont {Hasegawa},\ and\ \citenamefont
  {Fukamichi}}]{Fujita2003}%
  \BibitemOpen
  \bibfield  {author} {\bibinfo {author} {\bibfnamefont {A.}~\bibnamefont
  {Fujita}}, \bibinfo {author} {\bibfnamefont {S.}~\bibnamefont {Fujieda}},
  \bibinfo {author} {\bibfnamefont {Y.}~\bibnamefont {Hasegawa}}, \ and\
  \bibinfo {author} {\bibfnamefont {K.}~\bibnamefont {Fukamichi}},\ }\href
  {\doibase 10.1103/PhysRevB.67.104416} {\bibfield  {journal} {\bibinfo
  {journal} {Phys. Rev. B}\ }\textbf {\bibinfo {volume} {67}},\ \bibinfo
  {pages} {104416} (\bibinfo {year} {2003})}\BibitemShut {NoStop}%
\bibitem [{\citenamefont {Dung}\ \emph {et~al.}(2011)\citenamefont {Dung},
  \citenamefont {Ou}, \citenamefont {Caron}, \citenamefont {Zhang},
  \citenamefont {Thanh}, \citenamefont {de~Wijs}, \citenamefont {de~Groot},
  \citenamefont {Buschow},\ and\ \citenamefont {Br\"{u}ck}}]{Dung-Mixed}%
  \BibitemOpen
  \bibfield  {author} {\bibinfo {author} {\bibfnamefont {N.~H.}\ \bibnamefont
  {Dung}}, \bibinfo {author} {\bibfnamefont {Z.~Q.}\ \bibnamefont {Ou}},
  \bibinfo {author} {\bibfnamefont {L.}~\bibnamefont {Caron}}, \bibinfo
  {author} {\bibfnamefont {L.}~\bibnamefont {Zhang}}, \bibinfo {author}
  {\bibfnamefont {D.~T.~C.}\ \bibnamefont {Thanh}}, \bibinfo {author}
  {\bibfnamefont {G.~A.}\ \bibnamefont {de~Wijs}}, \bibinfo {author}
  {\bibfnamefont {R.~A.}\ \bibnamefont {de~Groot}}, \bibinfo {author}
  {\bibfnamefont {K.~H.~J.}\ \bibnamefont {Buschow}}, \ and\ \bibinfo {author}
  {\bibfnamefont {E.}~\bibnamefont {Br\"{u}ck}},\ }\href {\doibase
  10.1002/aenm.201100252} {\bibfield  {journal} {\bibinfo  {journal} {Adv.
  Energy Mater.}\ }\textbf {\bibinfo {volume} {1}},\ \bibinfo {pages} {1215}
  (\bibinfo {year} {2011})}\BibitemShut {NoStop}%
\bibitem [{\citenamefont {Wilkinson}, \citenamefont {Gingrich},\ and\
  \citenamefont {Shull}(1957)}]{wilkinson_magnetic_1957}%
  \BibitemOpen
  \bibfield  {author} {\bibinfo {author} {\bibfnamefont {M.~K.}\ \bibnamefont
  {Wilkinson}}, \bibinfo {author} {\bibfnamefont {N.~S.}\ \bibnamefont
  {Gingrich}}, \ and\ \bibinfo {author} {\bibfnamefont {C.~G.}\ \bibnamefont
  {Shull}},\ }\href {\doibase 10.1016/0022-3697(57)90074-4} {\bibfield
  {journal} {\bibinfo  {journal} {Journal of Physics and Chemistry of Solids}\
  }\textbf {\bibinfo {volume} {2}},\ \bibinfo {pages} {289} (\bibinfo {year}
  {1957})}\BibitemShut {NoStop}%
\bibitem [{\citenamefont {Cloud}\ \emph {et~al.}(1960)\citenamefont {Cloud},
  \citenamefont {Jarrett}, \citenamefont {Austin},\ and\ \citenamefont
  {Adelson}}]{cloud_neutron_1960}%
  \BibitemOpen
  \bibfield  {author} {\bibinfo {author} {\bibfnamefont {W.~H.}\ \bibnamefont
  {Cloud}}, \bibinfo {author} {\bibfnamefont {H.~S.}\ \bibnamefont {Jarrett}},
  \bibinfo {author} {\bibfnamefont {A.~E.}\ \bibnamefont {Austin}}, \ and\
  \bibinfo {author} {\bibfnamefont {E.}~\bibnamefont {Adelson}},\ }\href
  {\doibase 10.1103/PhysRev.120.1969} {\bibfield  {journal} {\bibinfo
  {journal} {Physical Review}\ }\textbf {\bibinfo {volume} {120}},\ \bibinfo
  {pages} {1969} (\bibinfo {year} {1960})}\BibitemShut {NoStop}%
\bibitem [{\citenamefont {Swoboda}\ \emph {et~al.}(1960)\citenamefont
  {Swoboda}, \citenamefont {Cloud}, \citenamefont {Bither}, \citenamefont
  {Sadler},\ and\ \citenamefont {Jarrett}}]{swoboda_evidence_1960}%
  \BibitemOpen
  \bibfield  {author} {\bibinfo {author} {\bibfnamefont {T.~J.}\ \bibnamefont
  {Swoboda}}, \bibinfo {author} {\bibfnamefont {W.~H.}\ \bibnamefont {Cloud}},
  \bibinfo {author} {\bibfnamefont {T.~A.}\ \bibnamefont {Bither}}, \bibinfo
  {author} {\bibfnamefont {M.~S.}\ \bibnamefont {Sadler}}, \ and\ \bibinfo
  {author} {\bibfnamefont {H.~S.}\ \bibnamefont {Jarrett}},\ }\href@noop {}
  {\bibfield  {journal} {\bibinfo  {journal} {Physical Review Letters}\
  }\textbf {\bibinfo {volume} {4}},\ \bibinfo {pages} {509} (\bibinfo {year}
  {1960})}\BibitemShut {NoStop}%
\bibitem [{\citenamefont {Bither}\ \emph {et~al.}(1962)\citenamefont {Bither},
  \citenamefont {Walter}, \citenamefont {Cloud}, \citenamefont {Swoboda},\ and\
  \citenamefont {Bierstedt}}]{bither_new_1962}%
  \BibitemOpen
  \bibfield  {author} {\bibinfo {author} {\bibfnamefont {T.~A.}\ \bibnamefont
  {Bither}}, \bibinfo {author} {\bibfnamefont {P.~H.~L.}\ \bibnamefont
  {Walter}}, \bibinfo {author} {\bibfnamefont {W.~H.}\ \bibnamefont {Cloud}},
  \bibinfo {author} {\bibfnamefont {T.~J.}\ \bibnamefont {Swoboda}}, \ and\
  \bibinfo {author} {\bibfnamefont {P.~E.}\ \bibnamefont {Bierstedt}},\ }\href
  {\doibase 10.1063/1.1728723} {\bibfield  {journal} {\bibinfo  {journal}
  {Journal of Applied Physics}\ }\textbf {\bibinfo {volume} {33}},\ \bibinfo
  {pages} {1346} (\bibinfo {year} {1962})}\BibitemShut {NoStop}%
\bibitem [{\citenamefont {Flippen}\ and\ \citenamefont
  {Darnell}(1963)}]{flippen_entropy_1963}%
  \BibitemOpen
  \bibfield  {author} {\bibinfo {author} {\bibfnamefont {R.~B.}\ \bibnamefont
  {Flippen}}\ and\ \bibinfo {author} {\bibfnamefont {F.~J.}\ \bibnamefont
  {Darnell}},\ }\href {\doibase doi:10.1063/1.1729383} {\bibfield  {journal}
  {\bibinfo  {journal} {Journal of Applied Physics}\ }\textbf {\bibinfo
  {volume} {34}},\ \bibinfo {pages} {1094} (\bibinfo {year}
  {1963})}\BibitemShut {NoStop}%
\bibitem [{\citenamefont {Kanomata}\ and\ \citenamefont
  {Ido}(1984)}]{kanomata_magnetic_1984}%
  \BibitemOpen
  \bibfield  {author} {\bibinfo {author} {\bibfnamefont {T.}~\bibnamefont
  {Kanomata}}\ and\ \bibinfo {author} {\bibfnamefont {H.}~\bibnamefont {Ido}},\
  }\href {\doibase doi:10.1063/1.333558} {\bibfield  {journal} {\bibinfo
  {journal} {Journal of Applied Physics}\ }\textbf {\bibinfo {volume} {55}},\
  \bibinfo {pages} {2039} (\bibinfo {year} {1984})}\BibitemShut {NoStop}%
\bibitem [{\citenamefont {Zhang}\ and\ \citenamefont
  {Zhang}(2003)}]{zhang_metamagnetic-transition-induced_2003}%
  \BibitemOpen
  \bibfield  {author} {\bibinfo {author} {\bibfnamefont {Y.}~\bibnamefont
  {Zhang}}\ and\ \bibinfo {author} {\bibfnamefont {Z.}~\bibnamefont {Zhang}},\
  }\href {\doibase 10.1103/PhysRevB.67.132405} {\bibfield  {journal} {\bibinfo
  {journal} {Physical Review B}\ }\textbf {\bibinfo {volume} {67}},\ \bibinfo
  {pages} {132405} (\bibinfo {year} {2003})}\BibitemShut {NoStop}%
\bibitem [{\citenamefont {Ohshima}, \citenamefont {Wakiyama},\ and\
  \citenamefont {Anayama}(1979)}]{ohshima_magnetic_1979}%
  \BibitemOpen
  \bibfield  {author} {\bibinfo {author} {\bibfnamefont {S.}~\bibnamefont
  {Ohshima}}, \bibinfo {author} {\bibfnamefont {K.~F.}\ \bibnamefont
  {Wakiyama}}, \ and\ \bibinfo {author} {\bibfnamefont {T.}~\bibnamefont
  {Anayama}},\ }\href {\doibase 10.1143/JJAP.18.707} {\bibfield  {journal}
  {\bibinfo  {journal} {Japanese Journal of Applied Physics}\ }\textbf
  {\bibinfo {volume} {18}},\ \bibinfo {pages} {707} (\bibinfo {year}
  {1979})}\BibitemShut {NoStop}%
\bibitem [{\citenamefont {Rodr\'{i}guez-Carvajal}(1990)}]{FullProf}%
  \BibitemOpen
  \bibfield  {author} {\bibinfo {author} {\bibfnamefont {J.}~\bibnamefont
  {Rodr\'{i}guez-Carvajal}},\ }\href@noop {} {\bibfield  {journal} {\bibinfo
  {journal} {Satellite Meeting on Powder Diffraction of the XV IUCr Congress}\
  }\textbf {\bibinfo {volume} {127}} (\bibinfo {year} {1990})}\BibitemShut
  {NoStop}%
\bibitem [{\citenamefont {Rietveld}(1969)}]{rietveld}%
  \BibitemOpen
  \bibfield  {author} {\bibinfo {author} {\bibfnamefont {H.~M.}\ \bibnamefont
  {Rietveld}},\ }\href {\doibase 10.1107/S0021889869006558} {\bibfield
  {journal} {\bibinfo  {journal} {J. of Appl. Crystallogr.}\ }\textbf {\bibinfo
  {volume} {2}},\ \bibinfo {pages} {65} (\bibinfo {year} {1969})}\BibitemShut
  {NoStop}%
\bibitem [{\citenamefont {Caron}\ \emph {et~al.}(2009)\citenamefont {Caron},
  \citenamefont {Ou}, \citenamefont {Nguyen}, \citenamefont {Cam~Thanh},
  \citenamefont {Tegus},\ and\ \citenamefont {Br\"uck}}]{caron_2009}%
  \BibitemOpen
  \bibfield  {author} {\bibinfo {author} {\bibfnamefont {L.}~\bibnamefont
  {Caron}}, \bibinfo {author} {\bibfnamefont {Z.}~\bibnamefont {Ou}}, \bibinfo
  {author} {\bibfnamefont {T.}~\bibnamefont {Nguyen}}, \bibinfo {author}
  {\bibfnamefont {D.}~\bibnamefont {Cam~Thanh}}, \bibinfo {author}
  {\bibfnamefont {O.}~\bibnamefont {Tegus}}, \ and\ \bibinfo {author}
  {\bibfnamefont {E.}~\bibnamefont {Br\"uck}},\ }\href {\doibase
  10.1016/j.jmmm.2009.06.086} {\bibfield  {journal} {\bibinfo  {journal} {J.
  Magn. Magn. Mater.}\ }\textbf {\bibinfo {volume} {321}},\ \bibinfo {pages}
  {3559} (\bibinfo {year} {2009})}\BibitemShut {NoStop}%
\bibitem [{\citenamefont {Klaasse}\ and\ \citenamefont
  {Br\"{u}ck}(2008)}]{klaasse_heat-capacity_2008}%
  \BibitemOpen
  \bibfield  {author} {\bibinfo {author} {\bibfnamefont {J.~C.~P.}\
  \bibnamefont {Klaasse}}\ and\ \bibinfo {author} {\bibfnamefont {E.~H.}\
  \bibnamefont {Br\"{u}ck}},\ }\href {\doibase 10.1063/1.3043430} {\bibfield
  {journal} {\bibinfo  {journal} {Review of Scientific Instruments}\ }\textbf
  {\bibinfo {volume} {79}},\ \bibinfo {pages} {123906} (\bibinfo {year}
  {2008})}\BibitemShut {NoStop}%
\bibitem [{\citenamefont {Okita}\ and\ \citenamefont
  {Makino}(1968)}]{okita_crystal_1968}%
  \BibitemOpen
  \bibfield  {author} {\bibinfo {author} {\bibfnamefont {T.}~\bibnamefont
  {Okita}}\ and\ \bibinfo {author} {\bibfnamefont {Y.}~\bibnamefont {Makino}},\
  }\href {\doibase 10.1143/JPSJ.25.120} {\bibfield  {journal} {\bibinfo
  {journal} {Journal of the Physical Society of Japan}\ }\textbf {\bibinfo
  {volume} {25}},\ \bibinfo {pages} {120} (\bibinfo {year} {1968})}\BibitemShut
  {NoStop}%
\bibitem [{\citenamefont {Zou}\ \emph {et~al.}(2009)\citenamefont {Zou},
  \citenamefont {Shen}, \citenamefont {Gao}, \citenamefont {Shen},\ and\
  \citenamefont {Sun}}]{ADMA:ADMA200901435}%
  \BibitemOpen
  \bibfield  {author} {\bibinfo {author} {\bibfnamefont {J.-D.}\ \bibnamefont
  {Zou}}, \bibinfo {author} {\bibfnamefont {B.-G.}\ \bibnamefont {Shen}},
  \bibinfo {author} {\bibfnamefont {B.}~\bibnamefont {Gao}}, \bibinfo {author}
  {\bibfnamefont {J.}~\bibnamefont {Shen}}, \ and\ \bibinfo {author}
  {\bibfnamefont {J.-R.}\ \bibnamefont {Sun}},\ }\href {\doibase
  10.1002/adma.200901435} {\bibfield  {journal} {\bibinfo  {journal} {Advanced
  Materials}\ }\textbf {\bibinfo {volume} {21}},\ \bibinfo {pages} {3727}
  (\bibinfo {year} {2009})}\BibitemShut {NoStop}%
\bibitem [{\citenamefont {Sandeman}(2012)}]{sandeman_magnetocaloric_2012}%
  \BibitemOpen
  \bibfield  {author} {\bibinfo {author} {\bibfnamefont {K.~G.}\ \bibnamefont
  {Sandeman}},\ }\href {\doibase 10.1016/j.scriptamat.2012.02.045} {\bibfield
  {journal} {\bibinfo  {journal} {Scripta Materialia}\ }\textbf {\bibinfo
  {volume} {67}},\ \bibinfo {pages} {566} (\bibinfo {year} {2012})}\BibitemShut
  {NoStop}%
\end{thebibliography}
%

\end{document}